\documentclass[12pt]{iopart}
\pdfoutput=1
\usepackage{graphicx}
\usepackage[utf8]{inputenc} 
\usepackage{iopams}
\usepackage{array}
\usepackage{color}
\usepackage{algorithm}
\usepackage{algorithmicx}
\usepackage{algpseudocode}
\usepackage{microtype}
\usepackage{hyperref}
\usepackage[defaultlines=3,all]{nowidow}

\newcommand{\D}{\Delta_i}
\newcommand{\A}{A_{ji}}
\newcommand{\W}{W_{ji}}

\begin{document}

\title{Decoupling approximation robustly reconstructs directed dynamical networks}

\author{Nikola Simidjievski$^1\ddagger$, Jovan Tanevski$^1$\footnote{These authors contributed equally to this work.}, Bernard Ženko$^1$, Zoran Levnajić$^{2,1}$, Ljupčo Todorovski$^{3,1}$ and Sašo Džeroski$^1$}

\address{$^1$ Department of Knowledge Technologies, Jožef Stefan Institute, Ljubljana, Slovenia}
\address{$^2$ Complex systems and Data science Lab, Faculty of information studies in Novo mesto, Novo mesto, Slovenia}
\address{$^3$ Faculty of administration, University of Ljubljana, Ljubljana, Slovenia}
\ead{zoran.levnajic@fis.unm.si}

\begin{abstract}
Methods for reconstructing the topology of complex networks from time-resolved observations of node dynamics are gaining relevance across scientific disciplines. Of biggest practical interest are methods that make no assumptions about the properties of the dynamics, and can cope with noisy, short and incomplete trajectories. Ideal reconstruction in such scenario requires an exhaustive approach of simulating the dynamics for all possible network configurations and matching the simulated against the actual trajectories, which of course is computationally too costly for any realistic application. Relying on insights from equation discovery and machine learning, we here introduce \textit{decoupling approximation} of dynamical networks and propose a new reconstruction method based on it. Decoupling approximation consists of matching the simulated against the actual trajectories for each node individually rather than for the entire network at once. Despite drastic reduction of the computational cost that this approximation entails, we find our method's performance to be very close to that of the ideal method. In particular, we not only make no assumptions about the properties of the trajectories, but provide strong evidence that our methods' performance is largely independent of the dynamical regime at hand. Of crucial relevance for practical applications, we also find our method to be extremely robust to both length and resolution of the trajectories and relatively insensitive to noise.
\end{abstract}

\noindent \emph{Accepted at} \NJP \url{https://dx.doi.org/10.1088/1367-2630/aae941}


\maketitle

\section{Introduction}

Complex networks are nowadays a standard paradigm for representing complex systems. Powered by insights from graph theory and dynamical systems, this paradigm has allowed for unprecedented improvement of our grasp over complex systems in nature, society and technology~\cite{newman,costa,estrada,easley,alb,pastor}. The functioning of a real complex network is a joint effect of its topology (structure) and its dynamics (interactions). The former refers to the connection patterns among its nodes (units)~\cite{stefano,hebert} and the latter to the ways in which these nodes interact among them~\cite{alex,mason}.

The foremost challenge in this vibrant field is the inverse problem of reconstructing (or inferring) the network topology from observations of the dynamics of its nodes. Namely, while the dynamical behavior of some or all network nodes can often be measured, structural details of many real networks remain elusive. Yet knowing how the topologies of real networks are made is crucial for our understanding of their functioning, especially in cases of more intricate structures such as networks of networks~\cite{dror}. In addition, elucidating architectural patterns of real network has its practical dimension that includes control of complex networks~\cite{ruths} and design of networks with prescribed functions~\cite{ja-sam}. However, the problem of network reconstruction is far from trivial given the issue of observability of complex systems~\cite{liu}. Therefore, the development of efficient methods of network reconstruction is vital, and it amounts to solving the inverse problem of estimating the presence/absence of links between pairs of nodes based on time-resolved measurements of their dynamics.

Actually, network reconstruction is becoming a field of its own within network science~\cite{timme1}. It brings together methodological disciplines such as computer science and statistics with domain sciences such as physics, sociology, biology and neuroscience. Within the context of physics, a myriad of methods have been proposed over the past decade. They are typically anchored in physical insights about network's collective/emergent dynamics~\cite{timme2,timme3}. This primarily includes synchronization as the best researched paradigm of collective dynamics~\cite{alex}, in both its theoretical~\cite{albert,luce} and experimental aspect~\cite{blaha,kralemann}. Methods applicable to sparse data have been developed~\cite{han}, as well as methods that work in the presence of noise~\cite{stankovski1}, or out of equilibrium~\cite{yasser}. Invasive methods assume that one is able to interfere with the system and extract the information from transients~\cite{us2}. Other methods use compressive sensing~\cite{sensing,grebogi} or elaborate statistics related to derivative-variable correlations~\cite{us1,marc}. Another set of methods attempts to grasp the situations relevant for inferring networks of neurons~\cite{rok,arkady}, or even social networks based on infection statistics~\cite{kocarev}. On the other hand, somewhat less effort has been invested in the development of methods that can reconstruct the interaction (coupling) functions and not necessarily the network topology~\cite{stankovski2}.

The problem of reconstructing the network from dynamical data is not to be confused with the problem of link prediction or network completion~\cite{roger,jure,tiago}. The latter refers to assessing the existence of a link by extracting the patterns of connectivity in the surrounding network. This approach does not involve dynamics on the nodes, but it is useful in completing the topologies of real networks that are often noisy due to experimental limitations. In this paper we will be dealing exclusively with the former model, where the entire topology of the studied network is hidden in a "black box" and is reconstructed based on observations of its node dynamics.

On a different front, in the context of computer science, the discipline of machine learning has been blossoming over the past decades~\cite{pat,vapnik,hastie}. Central to machine learning is the development of algorithms that are able to extract patterns and information from a set of observations and use them to make reliable models and predictions\footnote[7]{"to learn" from data in the context of machine learning means to extract various features from the dataset by performing different analyses. By making a Fourier decomposition of a time signal, physicist "learns" about ratio of various harmonics in the signal. By looking at e.g. statistics of specific words in an email, computer scientist can identify it as genuine or spam.}. Along these lines, physicists have over the past decade recognized that the remarkable ability of machine learning to classify and characterize complex sets of data can be useful in physics as well. For example, condensed matter physics is notoriously faced with problems where the size of the state space grows exponentially with the number of particles, which is reminiscent of the 'curse of dimensionality', well-known in machine learning. By knowing how to treat this curse, algorithms able to identify phases of matter and transitions between them were designed, including non-trivial states without conventional order parameter~\cite{phases1,phases2}. Algorithms able to recognize polymer structures are now available~\cite{qianshi}, along with algorithms that can identify particles in glassy systems susceptible to rearrangements~\cite{schoenholz}, or predict the physical properties of various compounds~\cite{seko}. Similar approaches were used in quantum computing~\cite{hentschel,yi}, to find density functionals~\cite{snyder}, and in the context of general inference problems in physics~\cite{toussaint}, including complex networks~\cite{zanin} and non-linear dynamics~\cite{wen}. It has in fact been claimed that machine learning "may soon become as common in physics as numerical simulations or calculus"~\cite{lenka}.

Coming back to the problem of network reconstruction, machine learning has a long and successful history of building models of natural and social phenomena from the available data. In particular, \textit{equation discovery} is a field of machine learning devoted to studying and developing methods for automated discovery of quantitative laws and models, expressed in the form of equations, from knowledge and data~\cite{scidisc}. Through the years, the focus of equation discovery has shifted from reconstructing well-known quantitative laws from history of science~\cite{ed-laws} towards automated modeling of dynamic systems~\cite{ed-dynamics,bridewell}. The methods for equation discovery make use of search algorithms~\cite{dzeroski}, genetic programming~\cite{lipson} or sparse regression techniques~\cite{sparsereg} to identify the structure and parameters of differential equations that model the dynamics of the system under investigation. In contrast to inference methods developed from physical insights, machine learning does not rely on the collective or other empirical properties of the studied dynamics. Consequently, machine learning methods usually make no assumptions about the studied system and are hence robust also to dynamical properties, such as chaotic or periodic motion. While this often makes them computationally more demanding, it widens their applicability in real-world scenarios, including treating dynamical systems of physical interest. In particular, equation discovery methods have been successfully used to construct reliable models of population dynamics~\cite{nikola-plos}, disease spread \cite{jovan-sysbio} and gene regulatory networks \cite{jovan-scirep}.

Inspired by the above observations, we here propose a new method of reconstructing dynamical networks formulated in the physical context, relying on insights from equation discovery. In particular, we apply the state-of-the-art automated modeling framework \textit{ProBMoT}\footnote[7]{available at \url{http://probmot.ijs.si}} \cite{probmot,nectar2017}, and tackle the challenge of network reconstruction by inferring the topology of a dynamical network from trajectories (time series) measured at individual nodes. Our method is based on what we call \textit{decoupling approximation}: we seek to reconstruct the network topology examining the nodes individually. We formulate the problem as it is normally done in physics and adjust ProBMoT for this task. As we show in what follows, our method displays good robustness to (white) noise and excellent performance in cases of time series incompleteness or bad resolution. Our method also shows decent robustness to dynamical regimes (information content of the time series) and is able to extract useful information also from relatively poor dynamics: reconstruction from regular/periodic dynamics is only slightly less precise than reconstruction from highly informative time series.

The rest of this paper is organized as follows. In the next section we explain our reconstruction method employing physics terminology. The section Results is devoted to examining the performance of our method using a toy-model of a dynamical network with 20 nodes. In the final section we discuss our findings and scrutinize their limitations, providing guidelines for future research.


\section{Reconstruction method}

Here we explain in detail our reconstruction method. It is developed by extending the scope of ProBMoT framework for equation discovery \cite{probmot,nectar2017} and adjusting it for physically formulated problems. We begin by considering a complex dynamical network consisting of $N$ nodes. The network is directed (links are not symmetric), but is not weighted (all links have unit weights). Dynamical state of a node $i$ at time $t$ is described by the variable $x_i (t)$, with $i=1,..., N$. The system's dynamics (time-evolution) is defined by:
\begin{equation} 
\dot x_i = \sum_{j=1}^N  \A f (x_j)  \, , 
\label{eq-1} 
\end{equation}
where the non-symmetric adjacency matrix $\A$ of dimensions $N \times N$ reports whether the node $j$ influences the node $i$ ($\A = 1$), or not ($\A = 0$). Function $f$ models the way that nodes interact among them (which is of course relevant only if the corresponding matrix element $\A$ is non-zero). Function $f$ is the same for all links. Thus, the dynamics of the node $i$ is a cumulative effect of interactions coming from the neighbouring nodes. This is a standard model of complex dynamical networks, capturing the properties of many systems of physical interest.

We make the following three assumptions regarding the information that is available about the system under investigation: 

\begin{itemize}
\item dynamical system (network) evolves according to Equation~\ref{eq-1}, 
\item mathematical form of interaction (coupling) function $f$ is known, 
\item a discrete trajectory consisting of $L$ values $x_i(t_1), \ldots , x_i(t_L)$ (time series) is available for each node $i$. 
\end{itemize}

We thus have at our disposal $N$ time series, each of length $L$ (in a realistic scenario these time series would come from an empirical measurement). The measurements of $x_i$ are separated by a uniform observation interval $\delta_t$ defining the resolution. For simplicity, we also assume that our network has no self-loops ($A_{ii}=0$ for each $i$), although this assumption is not crucial. Our goal now is to reconstruct (infer) the matrix $\A$ starting from the above three assumptions.

We intend to develop a general method that makes no assumptions whatsoever on the properties of the dynamics itself (e.g. periodicity). To that end we proceed as follows. There are $N(N-1)$ possible (directed) links in our network (we assume no self-loops) between each pair of nodes $i$ and $j$. Each of them can either exist (link, $\A=1$) or not exists (we call this situation non-link, $\A=0$). That is to say, there are $2^{N(N-1)}$ possible network configurations, each characterized by a different combination of links and non-links. Since we have time series (trajectories) at our disposal, an immediate naive approach would be to take the first time-point $x_i(t_1)$ for each node $i$ as the initial condition and run the dynamics of Equation~\ref{eq-1} for each possible network configuration. Then, we could find the network configuration that leads to the best match between the simulated time series and the original time series. This network configuration would then be our reconstructed network. This is the ideal approach, which would guarantee excellent results. However, such an exhaustive combinatorial search of $O(2^{N(N-1)})$ is computationally intractable even for a network of modest size $N \approx 10$, which would forever limit any realistic use of such a method. What we need instead is an approach that would be as close as possible to the above, but with more acceptable computational costs.

To this aim we use insights from equation discovery. We begin by noting that the dynamics of any node $i$ is chiefly governed by inputs coming via in-links from other nodes, i.e., by the dynamics of those nodes $j$ from which there is a directed link pointing to $i$. Nodes $j$ for which $\A=0$ do not influence the dynamics of $i$ directly, i.e. in the first approximation\footnote{To solve Equation~\ref{eq-1} for node $i$ precisely, we should really consider the entire coupled system and not just the nodes influencing node $i$ directly. Namely, nodes $j$ with $\A=0$ can still influence node $i$ indirectly, via some of the nodes that influence it directly. In other words, while the equation for the derivative of $x_i$ only includes the neighbors of $i$, the equations for the derivatives of the latter may contain other nodes. What we here mean by "approximation" is the fact that, when searching for the best in-link configuration for the node $i$, we run (simulate) the dynamics for node $i$ by considering only the observed values for its immediate neighbors (with links pointing towards $i$), and not their properly simulated values, which would require running (simulating) the dynamics of the entire coupled system from Equation~\ref{eq-1}.}.

Since $N-1$ nodes can have links pointing to $i$, there are $2^{N-1}$ possible in-link configurations for the node $i$ (for an "in-link configuration" we intend one specific combination of links and non-links pointing to the node $i$). With this in mind we look for the best in-link configuration by running the dynamics of the node $i$ for all in-link configurations. Note that here we do not run the dynamics of the whole system Equation~\ref{eq-1}, but \textit{only} the dynamics of the node $i$. This is computationally very simple, but it has to be done $2^{N-1}$ times (for each node $i$), which can be costly. For numerical integration we use CVODE package from the SUNDIALS suite~\cite{CohenHindmarsh}, which relies on a multi-step variable-coefficient method and is suitable for a wide class of ODEs. This procedure leaves us with $2^{N-1}$ simulated time series for the node $i$, each corresponding to one in-link configuration. By choosing such sophisticated integrator we minimize the numerical integration errors, which basically makes all errors attributable to the suitability of the individual in-link configurations. We perform the same calculation for all network nodes in analogous fashion, which makes the total computational costs of this procedure $O(N2^{N-1})$. As we show in what follows, this approach offers a considerable reduction of computational cost while providing almost excellent reconstruction results. Other algorithmic details of above procedure are in Appendix A.

Next we consider the node $i$ and look for the in-link configuration most suitable for it. To do so we need a measure of discrepancy between an original time series $x_i(t_k)$ and a simulated one, which we commonly denote as $\hat{x_i}(t_k)$. We use the standard root mean squared error $\D$ defined as follows: 
\begin{equation}
\D = \D (\hat{x_i},x_i) = \sqrt{\frac{1}{L}\sum_{k=1}^{L} \Big( \hat{x_i}(t_k) - x_i(t_k) \Big)^2} \;.
\label{eqrmse}
\end{equation}
While the in-link configuration corresponding to the minimal value of $\D$ is the best candidate, we note that in-link configurations with slightly larger values of $\D$ are also of interest. Due to the reality of empirical measurements, it is conceivable that many in-link configurations will display similarly small values of $\D$. This indicates that our best candidate is to be extracted from several among the best in-link configurations.

\begin{figure}
\centering
\includegraphics[width=.9\textwidth]{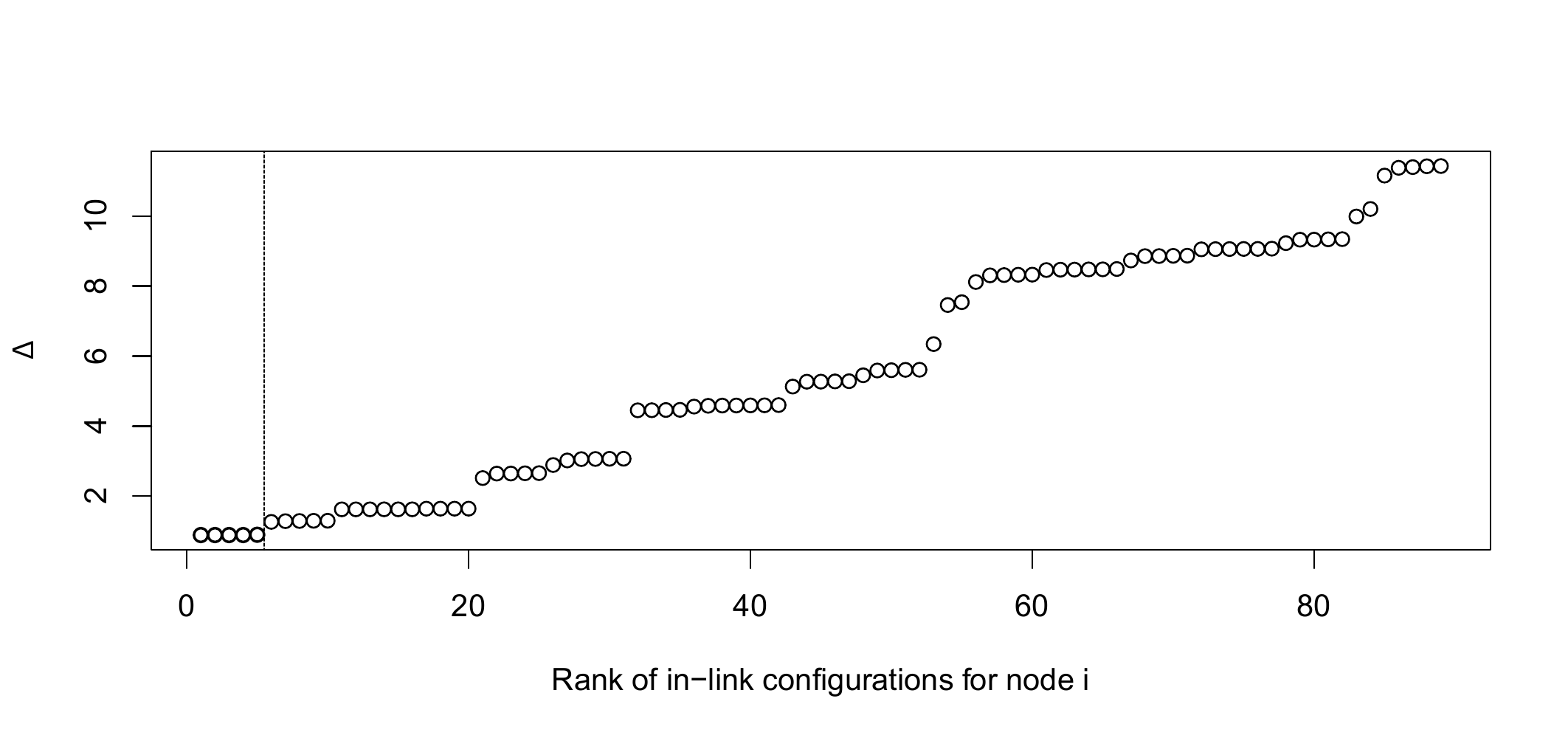}
\caption{Example of ranking of in-link configurations for a single node $i$ and definition of the corresponding plateau. Each dot represents an in-link configuration and displays its value of $\D$ on $y$ axis. In-link configurations are arranged to form a non-decreasing sequence of $\D$ values, which defines the rank of in-link configurations reported on $x$ axis. The in-link configurations form the plateau and are separated from the remaining in-link configurations by the vertical line. Only the in-link configurations left of the vertical (dashed) line are considered for calculating the propensity matrix $W$. In-link configurations to the right of the vertical line are discarded. This example of ranking was derived from one of the trajectories considered later in the paper (next Section).}
\label{figranking}
\end{figure}

We proceed by ranking the in-link configurations for each node $i$ according to their respective values of $\D$. For each node $i$ we consider the plateau of top-ranked in-link configurations with indistinguishable $\D$. We take two consecutive in-link configurations to be indistinguishable if the relative difference of their values of $\D$ is less or equal to 10\%. The optimal value for this percentage is to be set depending on each particular case (for system here studied, we found that 10\% yields the smallest possible number of "equally" performing configurations while achieving the highest AUC, which start to deteriorate as percentage is increased). In Figure~\ref{figranking} we illustrate the ranking of in-link configurations and the construction of the plateau. In-link configurations not belonging to the plateau are not of further interest. 

Now, looking at another node $j$ that may or may not have a directed link pointing towards $i$, we note that some in-link configurations within the plateau will assume the existence of such a link, and others will not. We define the \textit{propensity} (likelihood) of directed link $j \rightarrow i$ as the fraction of in-link configurations within the plateau that do assume its existence. High (low) propensity means that very many (very few) in-link configurations within plateau assume the existence of that link. We arrange the propensity values for all directed links into the propensity matrix $W$, so that the propensity of link $j \rightarrow i$ corresponds to the matrix element $\W$. Once equipped with the propensity matrix $\W$, we define our reconstructed adjacency matrix $\hat{\A}$ as follows. We define the threshold $\theta$, which has a value between 0 and 1. Then, for a given $\theta$ and for all pairs of nodes $j$ and $i$ we define: 
\begin{equation}
    \left\{
        \begin{array}{lll}
        \hat{\A} &= 1 \;\;\; & \mbox{if} \;\;\; \W \geq \theta, \\
        \hat{\A} &= 0 \;\;\; & \mbox{otherwise}
        \end{array}
    \right.
\end{equation}
The threshold $\theta$ indicates how much propensity for link $j \rightarrow i$ we require to recognize $j \rightarrow i$ as a link in the reconstructed adjacency matrix $\hat{\A}$. Ideally, we would obtain all propensities equal to 1, which would immediately make $\hat{\A} = \W$. Realistically, many propensities will vary between 0 and 1, so setting $\theta$ high (low) means that we wish to recognize only (also) the links with strong (weak) propensity.

\begin{figure}
\centering
\includegraphics[width=.5\textwidth]{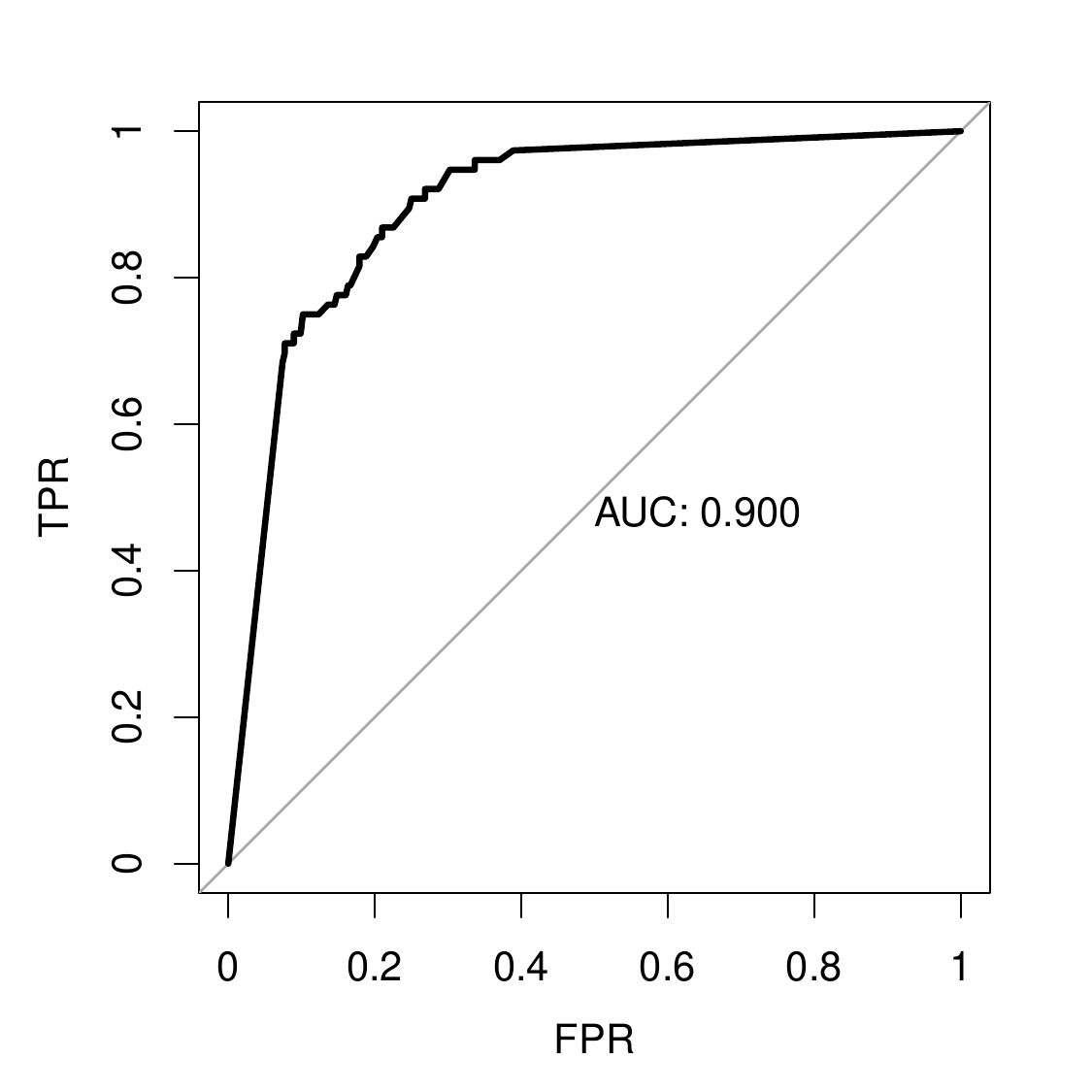}
\caption{Example of a ROC curve. By changing the value of the threshold $\theta$ we get different combinations of $TPR$ and $FPR$ values, which form the by black ROC curve. In essence, we wish the curve to initially grow as steeply as possible, reaching high $TPR$ while keeping $FPR$ low. Shown for comparison in grey is the diagonal line that corresponds to a fully non-discriminatory ROC curve. The AUC value for the black ROC curve is AUC=0.9, much better than the AUC=0.5 corresponding to the grey diagonal. This ROC example corresponds to one of the trajectories considered later in the paper (next Section).}
\label{figroc}
\end{figure}

The threshold $\theta$ also offers a robust way to quantify the quality (efficiency) of our reconstruction, i.e., the agreement between matrices $\hat{\A}$ and $\A$. One can examine how this agreement depends on the choice of $\theta$. Lower values of $\theta$ will allow to correctly detect many links (true positives), but may also lead to non-links incorrectly recognized as links (false positives). In contrast, higher values of $\theta$ may hide some of the actual links (false negatives), but will also correctly exclude many non-links (true negatives). The standard way to capture this relation is via \textit{receiver-operating characteristic curve} (ROC curve)~\cite{roc-ml}, in which by varying $\theta$ one plots the true positive rate ($TPR$) against the false positive rate ($FPR$). An illustration of a ROC curve is shown in Figure~\ref{figroc}. 

ROC curve details the trade-off of between $TPR$ and $FPR$ as $\theta$ changes. Ideally, we would get all true positives with no false positives, corresponding to the ROC curve that immediately reaches 1. The worst case scenario is a completely non-discriminatory ROC curve coinciding with the diagonal line ($TPR = FPR$, see Figure~\ref{figroc}), where our reconstruction is no better than random guessing. That means that area under the ROC curve (AUC) serves as an excellent indicator of the quality (efficiency) of reconstruction. AUC is formally defined as the integral of the ROC curve for the corresponding range of $\theta$, thus representing the probability that a randomly chosen actual link will be correctly recognized as a link by our reconstruction method. Values of AUC range from 0.5 (worst) to 1 (best), we will rely on them in reporting the performance of our reconstruction method (next Section).

To sum up the idea behind our method, we decouple the network dynamics by examining each node individually, thus reducing the computational cost of our method to the level suitable for moderate size networks. This of course is only a first approximation, since complete dynamics of a node ultimately depends not just on immediate neighbours. However, as we will show in what follows, our method displays excellent performance, is robust to different dynamical regimes and noise, and shows very slow deterioration with shortening of time series or worsening of time series resolution. Performance is even better when reconstructing the network from several time series corresponding to different initial conditions rather than just one. Note also that in this formulation, our method is applicable also to the more general case when the interaction function $f$ depends on both $x_j$ and $x_i$ (cf. RHS of Eq.~\ref{eq-1}). More on the algorithmic aspects of our method can be found in Appendix A.


\section{Results}

\begin{figure}[!htb]
\centering
\includegraphics[width=0.6\textwidth]{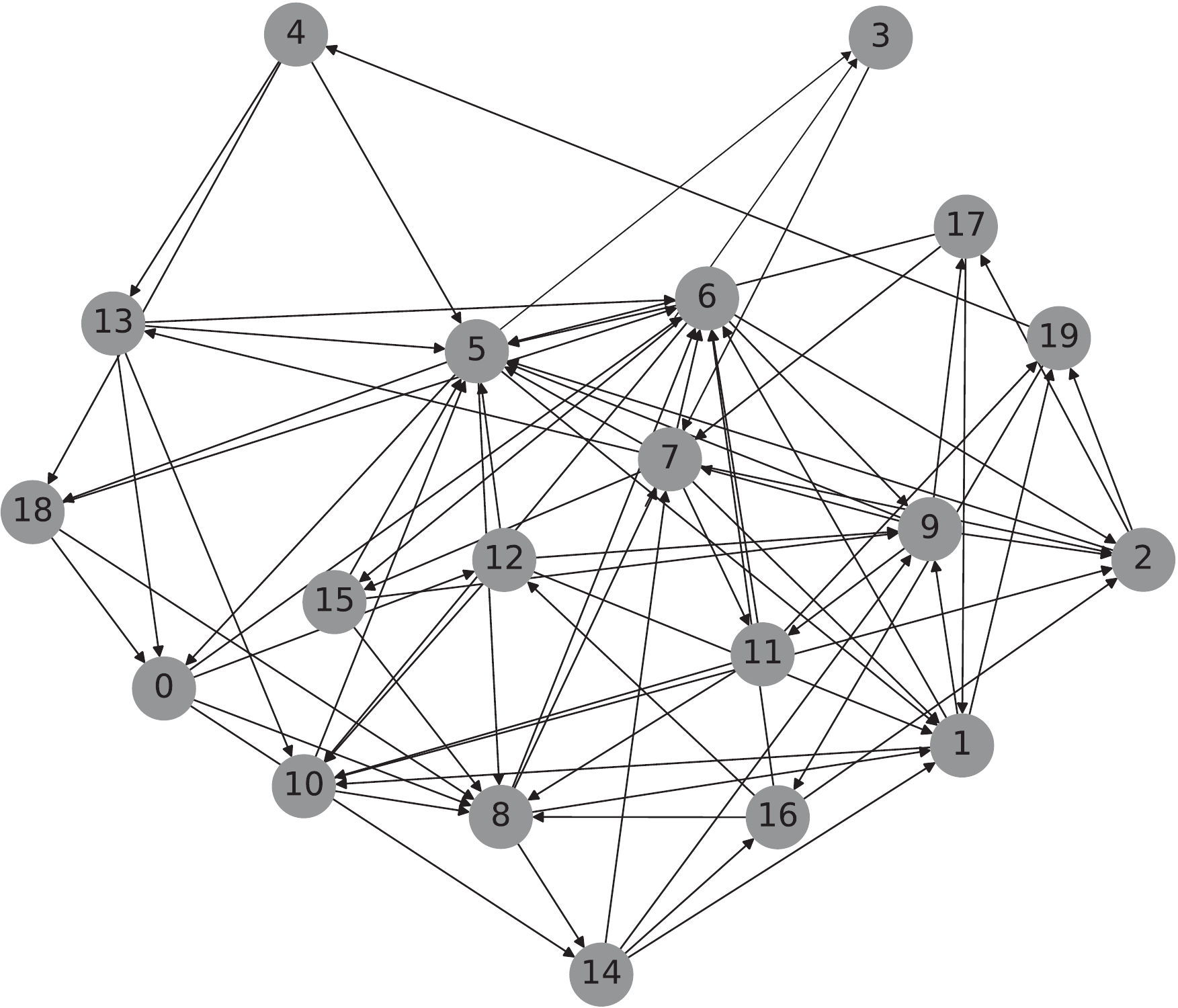}
\caption{The directed network used for testing the performance of our method. It contains $N=20$ nodes and 76 directed links, it is generated via standard preferential attachment procedure. We picked a random realization that best suits our needs in testing our reconstruction method.}
\label{fig-network}
\end{figure}

In this Section we test the performance of our method. To that end we design a toy-model dynamical network, suitable for testing. We take a directed network with $N=20$ nodes and 76 directed links (no self-loops). The directed links are placed between 76 pairs of nodes, which are chosen randomly via standard preferential attachment. We picked a network realization with nodes having very diverse in-degrees and out-degrees. Each node has at least one in-link and one out-link. The network structure is shown in Figure~\ref{fig-network}. 
For the interaction function $f$ (see Equation~\ref{eq-1}) we take $f = \tanh$. This corresponds to a well-known physical model of collective network dynamics~\cite{us1,hansel}, with very rich dynamics that is needed for testing the performance of our method against various dynamical behaviors. In particular, depending on the initial conditions, this system can exhibit diverse dynamical patterns, ranging from rich motion with high variability to poor dynamics with scarce phase space coverage.

We run the system Equation~\ref{eq-1} with $f = \tanh$ on network Figure~\ref{fig-network} and store the obtained trajectories. Each simulation starts from a set of initial conditions, chosen uniformly at random from $[-10, 10]$ for each node. For each initial condition, the system is numerically integrated from $t = 0$ to $t = 10$. During the run, $L = 100$ values for each node are stored, equally spaced with time-resolution $\delta_t = 0.1$. This leads to the time series $x_i (t_k)$ ($k$ goes from 1 to $L=100$) for each node ($i$ goes from 1 to $N=20$). That is to say, for each choice of initial conditions we obtain 20 time series (one for each node) of length 100 time points.

Of course, we cannot test the performance of our method on all possible system trajectories, meaning we need to make a selection of several trajectories and limit our analysis to them. Yet in order to test the impact of dynamical richness on the quality of reconstruction, we do need trajectories exhibiting diverse dynamical properties. We naturally expect that more complex trajectories that contain more information about the underlying network will lead to better reconstructions. It is often the case that quality of reconstruction strongly depends on the richness of the dynamics at hand. This, however, limits the applicability of such methods in scenarios of poor dynamics, which cannot be excluded in realistic situations. But as already stated, our goal here is to design a general reconstruction method that would be as robust as possible to the richness of the dynamics and able to cope with poor dynamics. With this in mind, we made a selection of five different trajectories corresponding to five different initial conditions. These trajectories are selected to reflect as best possible the diversity of dynamical patterns exhibited by our system in terms of phase space coverage and dynamical variability. The detailed justification of this selection is given in the Appendix B. In what follows we test the performance of our method against these five trajectories, considering also their noisy versions. For easier orientation we label them as $\mathbf{T}_1$,... $\mathbf{T}_5$.

\begin{figure}[!htb]
\centering
\includegraphics[width=\textwidth]{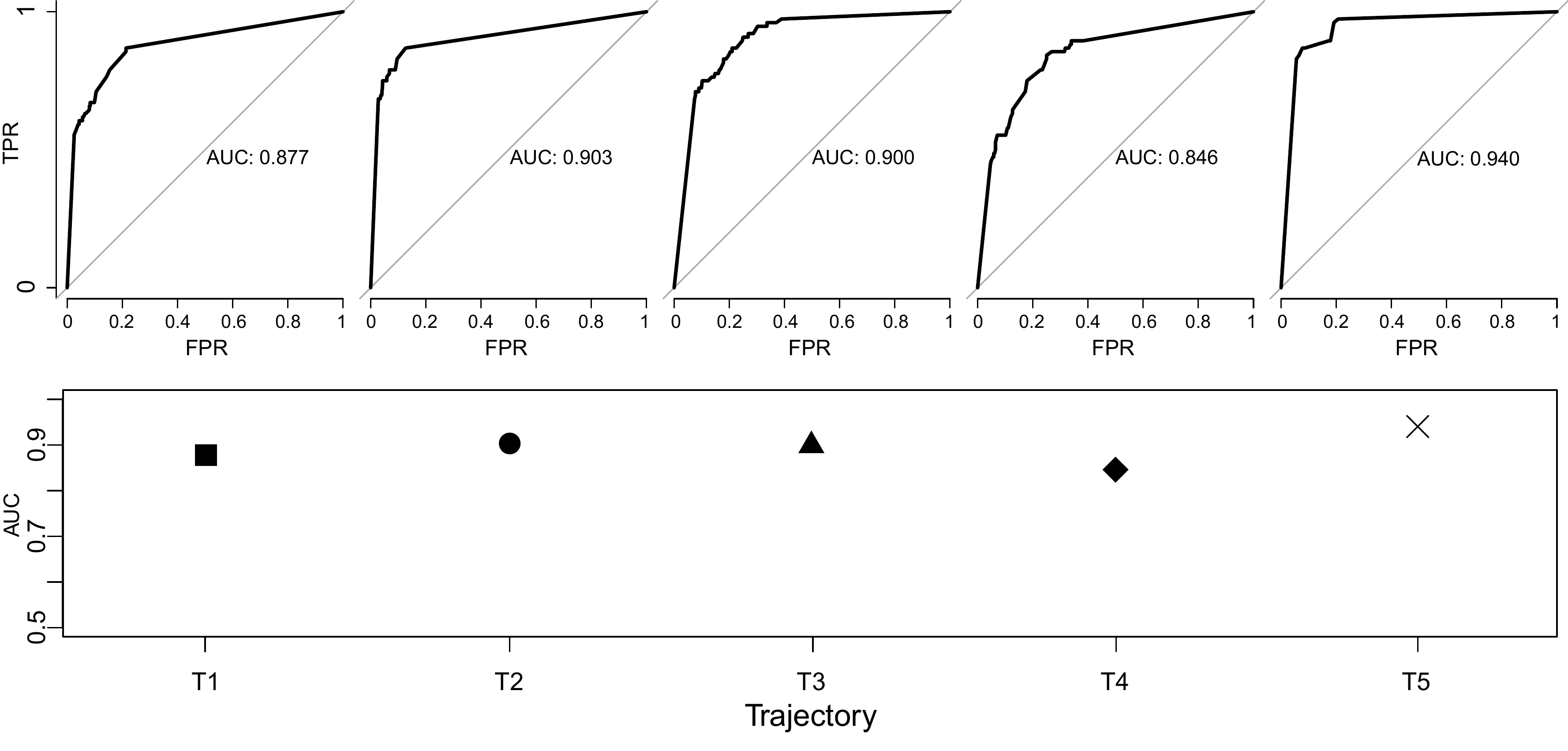}
\caption{Top panel: ROC curves for the five selected trajectories $\mathbf{T}_1$,... $\mathbf{T}_5$ obtained using our method described in the previous Section (cf. Figure~\ref{figroc}). Bottom panel: AUC values obtained by integrating these five ROC curves. Top and bottom plots are aligned for clarity.}
\label{figrocfive}
\end{figure}

Now we start testing the performance of our method by applying it to the five selected trajectories. For each trajectory we obtain a ROC curve shown in Figure~\ref{figrocfive} (top panel). ROC curves are slightly different from each other since they reflect the varying dynamical nature of the five trajectories. We compute the integral of each ROC curve and obtain the AUC values. We show the numerical AUC values in Figure~\ref{figrocfive} (bottom plot), which reflect the quality of reconstruction.
ROC curves and AUC values indicate very good reconstruction in all cases, with all AUC values around AUC = 0.9. Given the trajectory length of only $L=100$ points, this confirms that our method is able to offer useful reconstruction results, despite the approximation of decoupling the dynamics of individual nodes. The results also indicate that reconstruction depends very little on the dynamical properties of each trajectory, since the five values show no clear correlation with the apparent richness of dynamics visible in Figure~\ref{figtrajectories}. This suggests that our method is largely robust to dynamical regimes and time series complexity. It also suggests that the method can -- at least in principle -- extract information also from dynamically poor trajectories. In the Subsections that follow we test the robustness of our method's performance to three experimentally realistic scenarios with potentially detrimental influences: noise, shortening of time series length and degradation of time series resolution.

\subsection{Performance for noisy trajectories}
\begin{figure}[!htb]
\centering
\includegraphics[width=\textwidth]{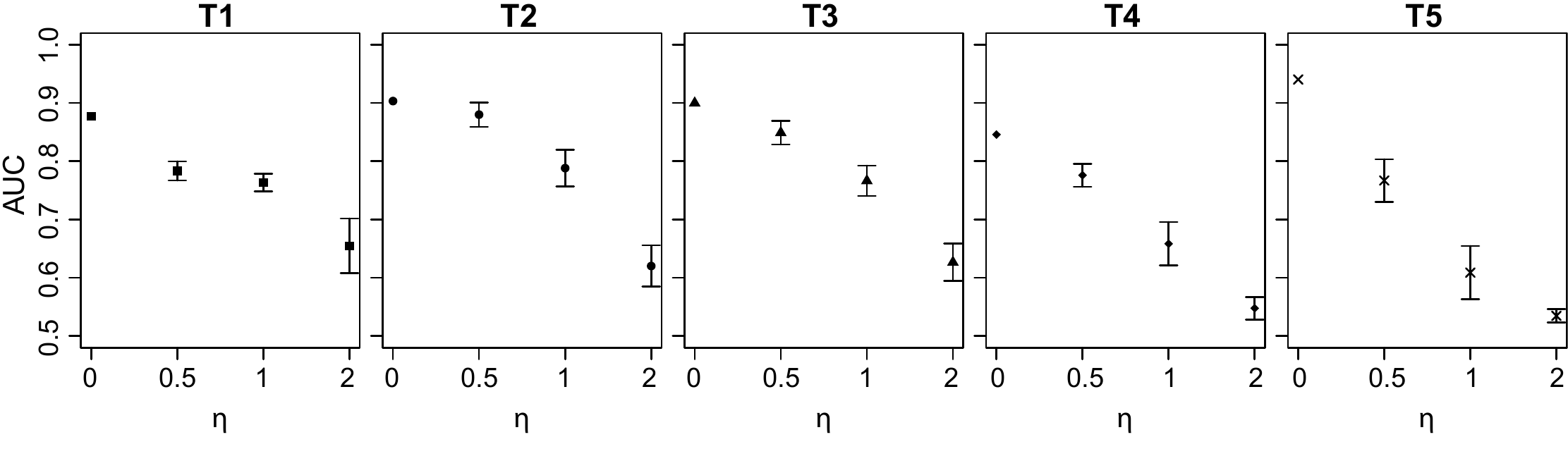}
\caption{AUC values for reconstruction from noisy trajectories. Each plot corresponds to one trajectory as indicated in the title. In each plot, noise strength $\eta$ is reported on $x$ axis (including the noise-free case $\eta=0$), while AUC values are shown on $y$ axis. Errorbars correspond to standard deviations calculated over 20 random realizations of noising.}
\label{fignoise}
\end{figure}

Noise is ubiquitous in all physical experiments and measurements. We thus test our method against the presence of noise in time series. To that end we (artificially) introduce uncorrelated noise in our five trajectories by adding to each data point of the original time series a random value that we pick uniformly from $[-\eta,\eta]$. We introduce three levels of noise strength, $\eta=0.5, 1.0, 2.0$. For each trajectory and each noise strength, we make 20 random realizations of noising (cf. Figure~\ref{figtrajectories} in Appendix B). Then we apply our method to all noisy trajectories and report the results, averaging the 20 AUC values obtained for different noise realizations for the same noise strength. In Figure~\ref{fignoise} we report the averaged AUC values for our five noisy trajectories. Each plot regards a single trajectory, with different noise strengths reported on the $x$ axis (including the noise-free case, $\eta=0$).
For each of the five plots (five trajectories) zero-noise level corresponds to what seen in Figure~\ref{figrocfive}. As the noise increases, we observe a gradual deterioration of the AUC values, which is expected for all trajectories. However, for weak noise, the AUC values are still close to the AUC values for noise-free case, and decay considerably only as the noise grows stronger. Considering that trajectories generally range between -10 and 10, noise level of 2.0 is in fact fairly strong (cf. Figure~\ref{figtrajectories}). We hence conclude that at least for weak noise our method shows very good performance. We also note that even in the case of strongest noise our method performs better than random guessing.

\subsection{Performance for varying trajectory length}
\begin{figure}[!htb]
\centering
\includegraphics[width=\textwidth]{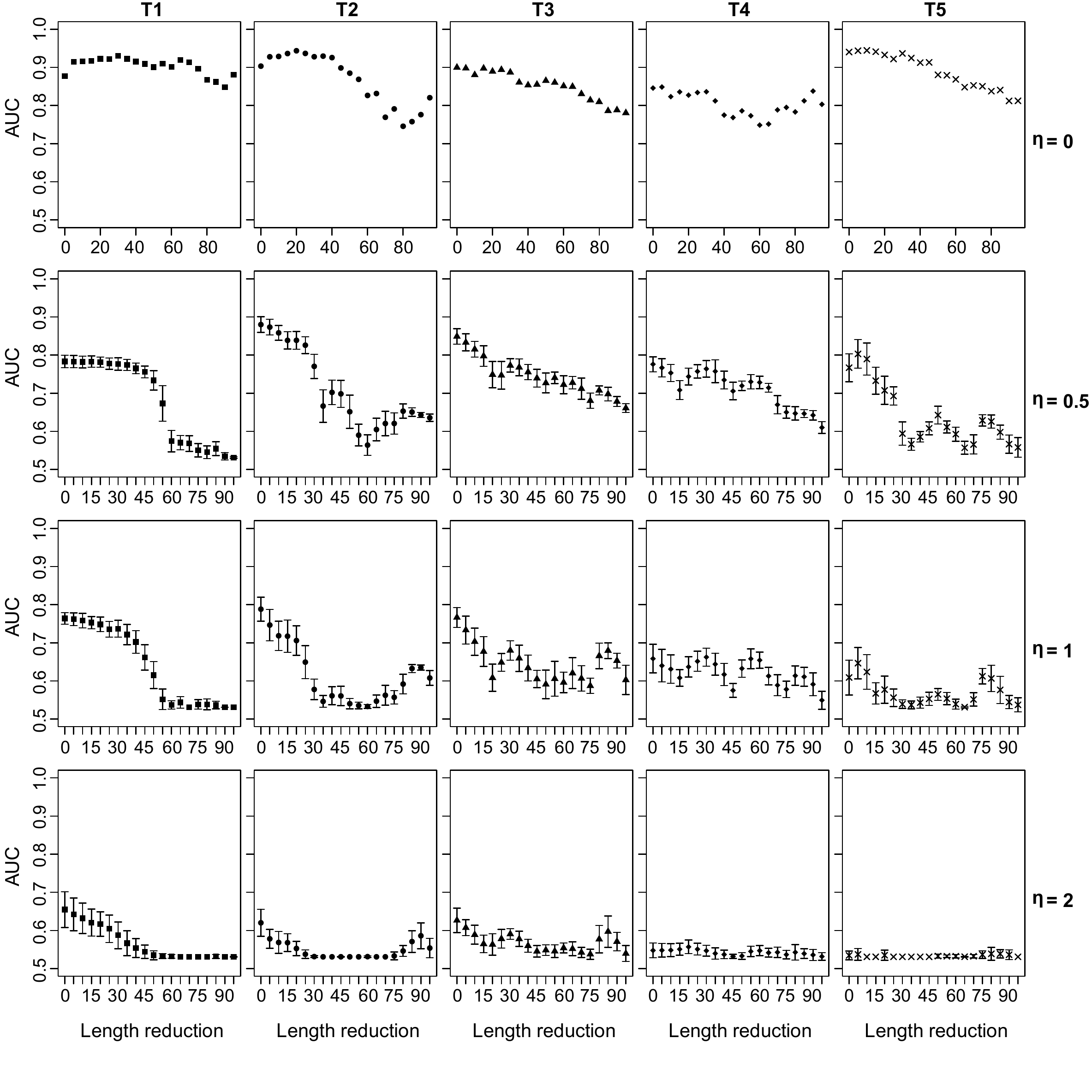}
\caption{AUC values for reconstruction from trajectories of variable length (truncated trajectories) with and without noise. Each column refers to one trajectory as indicated by the label at the top. Each row (panel) refers to one noise strength, $\eta=0$ (noise-free), $\eta=0.5$, $\eta=1.0$ and $\eta=2.0$, respectively (as indicated on the right). Truncation (reduction of time series length) is reported on $x$ axis, with 0 corresponding to full (non-truncated) trajectory. AUC values are reported on $y$ axis. Errorbars correspond to standard deviations calculated over 20 random realizations of noising.}
\label{truncation-noise}
\end{figure}

Real experimental measurements are difficult and often expensive, which is why real data often consists of only a few data points (measurements). With this in mind, we next examine how does our method respond to reducing the length of time series (i.e., truncating time series). To test this, we first go back to the original noise-free trajectories with 100 data points. Now, instead of considering the entire trajectory, we consider only a certain number of points starting from $x_i(t_1)$ and discard the rest. We hence cut time points from the end, step by step, 5 points in each truncation step. To each truncated time series we apply our method and obtain the respective AUC value. We report these AUC values in Figure~\ref{truncation-noise} (top panel), where each trajectory corresponds to one plot in which on the $x$ axis we show the number of removed points (non truncated trajectories correspond to length reduction 0). 
We find no dramatic worsening of reconstruction quality for any of the dynamical regimes. $\mathbf{T}_1$ and $\mathbf{T}_4$ show almost no overall worsening at all, while the other trajectories display only minimum worsening. This indicates that our method is extremely robust to time series length and able to extract useful information from very short time series. To be more precise: our method yields useful reconstruction even by utilizing only 5 data points (AUC roughly equal 0.8), which corresponds to the last AUC values in all plots. Since those 5 time points are separated by the original time resolution of $\delta_t = 0.1$, this means our method yields useful reconstruction even from poor phase space coverage.

Next we look at the noisy trajectories and examine the same sequence of truncations for all three levels of noise strength discussed earlier. The results are shown in three other panels in Figure~\ref{truncation-noise}, where each panel corresponds to one noise strength and each column to one trajectory. Unlike in the noise-free case, we here do observe gradual worsening of the method's performance due to truncation. As expected, worsening is faster when the noise is stronger. In the case of weakest noise, AUC values persist for considerable reduction of time series length. For stronger noise, AUC values deteriorate more rapidly. We note that for weak noise our method in general still yields reconstructions better than random guessing from only 5 time points.

\subsection{Performance for varying trajectory resolution}
\begin{figure}[!htb]
\centering
\includegraphics[width=\textwidth]{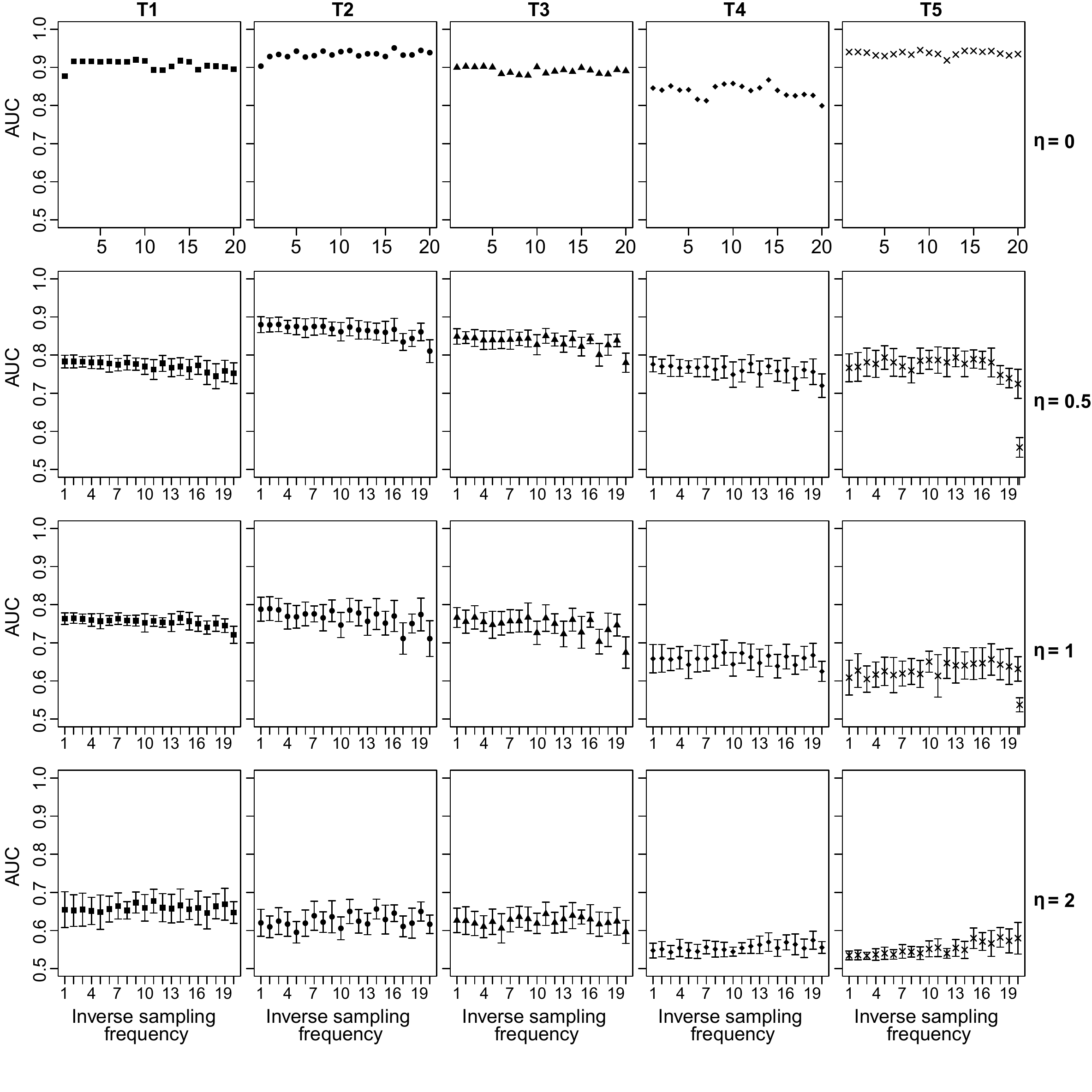}
\caption{AUC values for reconstruction from uniformly sampled trajectories with and without noise. Each column refers to one trajectory as indicated by the label on the top. Each row (panel) refers to one noise strength, $\eta=0$ (noise-free), $\eta=0.5$, $\eta=1.0$ and $\eta=2.0$, respectively (as indicated on the right). Inverse sampling frequency is reported on $x$ axis (1 means we consider all $L=100$ points, 20 means we consider every 20-$th$ point). AUC values are reported on $y$ axis. Errorbars correspond to standard deviations calculated over 20 random realizations of noising.}
\label{sampling-noise}
\end{figure}

Some experimental measurements can be only made with a limited time-resolution, which in practice means that we do not always have control over how many data points per unit time we obtain. This however can be critical, since the dynamical scale of the process under study can become shorter than the measurement resolution, obscuring the true nature of dynamics. To test the robustness of our method to time series resolution, we next repeat the testing process above, but this time by gradually reducing the time series resolution. That is to say, we sample each trajectory with a given frequency and consider only the sampled time points, discarding the rest. Again, we first start with noise-free trajectories and examine (inverse) sampling frequencies ranging from 1 (we consider all $L=100$ points) to 20 (we consider every 20-$th$ point, meaning only 5 points altogether, separated by $\delta_t = 2.0$). For each sampling frequency we apply our method and obtain the respective AUC value. In Figure~\ref{sampling-noise} (top panel) we report these results for each trajectory, with inverse sampling frequency reported on $x$ (full trajectory corresponds to 1). 
Remarkably, there is basically no worsening at all for any of the trajectories. AUC values for all sampling frequencies stay very close to the values for sampling frequency 1. This indicates that our method is very robust to the resolution of input data and able to extract useful information from time series with very bad resolution. Again, we underline that our method works from 5 time points practically with the same reconstruction quality as from 100 time points. While phase space coverage is good in this case, the resolution is not, hindering the estimation of derivatives needed for numerical integration. We hence attribute this robustness to both general reconstruction idea and the choice of numerical integrator. Of course, another pertinent issue is the relationship between the sampling frequency and the characteristic time scale of the dynamics. In fact, we expect that the rate of worsening of AUC values with the former will in general be related to the latter, which merits further investigation.

Next we carry out the same analysis but for noisy trajectories, again considering all three levels of noise strength, as done above in the Figure~\ref{truncation-noise}. The results are shown in three other panels in Figure~\ref{sampling-noise}, where again each panel corresponds to one noise strength and each column to one trajectory. Contrary to the case of time series truncation, here we see far less worsening due to noise. In fact, for weak noise, the values of AUC exhibit no visible deterioration with sampling frequency and stay very close to the noise-free case with the sampling rate 1. Even for strong noise the AUC values do not worsen much more than they would do otherwise, i.e., deterioration due to noise has much more influence than due to sampling. This appears to depend very little on the considered trajectory, although some trajectories show more AUC value fluctuations for different noise realization than others. And again, for weak noise our method in general still yields reconstructions better than random guessing for all trajectories even when using only 5 time points (every 20-$th$ point).

\subsection{Performance for multiple trajectories}
\begin{figure}
\centering
\includegraphics[width=.75\textwidth]{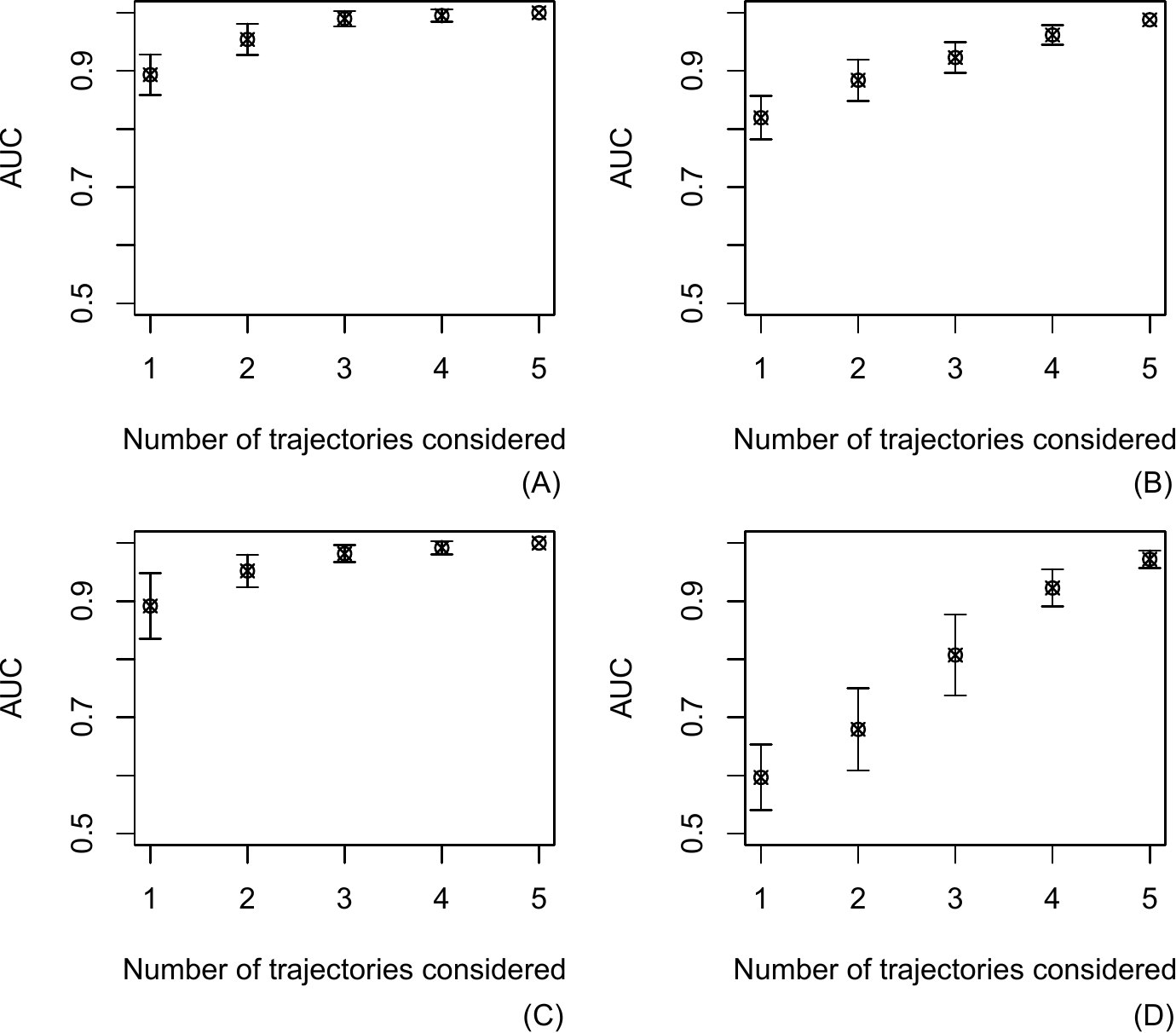}
\caption{Reconstruction from multiple trajectories. (A) Full length and full frequency. (B) Minimum length (5 points). (C) Minimum sampling (every 20-$th$ point), all without noise. (D) Full length and full frequency, maximum noise level $\eta=0.2$. In each plot, the number of trajectories is shown on $x$, while the AUC values are shown on $y$.  Errorbars correspond to standard deviations calculated over different combinations of trajectories.}
\label{multiple}
\end{figure}

So far, we examined the performance of our method when using only \textit{one} trajectory at a time, i.e., reconstructing the network from a single dynamical regime. Yet in practice (although rarely), we can be "lucky" and have access to multiple trajectories produced by the same system. One such scenario is when we can interfere with the system and reset its dynamics~\cite{us2}. Our method can be easily adjusted for reconstruction from multiple trajectories. Of course, we here expect much better performance, since not only we have more data to reconstruct from, but we also have dynamically more diverse data with better phase space coverage. To test our method in this scenario, we use it with 2, 3, 4 and all 5 trajectories. For cases of 2, 3 and 4 we consider all possible combinations of the trajectories. Results are reported in Figure~\ref{multiple}A, where on the $x$ axis we plot the number of trajectories used for reconstruction and on the $y$ axis the average AUC values obtained.
As expected, performance clearly improves when using more trajectories, reaching AUC=1 when 3 or more trajectories are used. While this was not as surprising, we now re-do this analysis but using shortened (truncated) and sampled trajectories. In particular, we use the maximal truncation (only 5 points taken from the beginning of the time series) and minimum sampling rate (sample only every 20-$th$ point). The results are shown in Figure~\ref{multiple}B and Figure~\ref{multiple}C respectively. The performance again improves drastically and with 5 trajectories reaches 1 (truncated) or a value very close to 1 (sampled). The sampling case in fact shows no major difference in performance as compared to the case of full trajectories. Finally, we examine the case of full trajectory length and resolution, but with strongest noise of $\eta=2.0$. Results are shown in Figure~\ref{multiple}D. Again, considering multiple trajectories clearly improves the performance. Note in fact that when all 5 trajectories are considered, the AUC practically reaches 1 also in this case. We conclude that using multiple trajectories clearly improves our method's performance, irrespective of which trajectories are used and irrespective of truncating, sampling or noising the trajectories.


\section{Discussion}

We designed and presented a new method of reconstructing (inferring) the topology of directed dynamical networks from time-resolved observations of their dynamics, assuming the knowledge of the dynamical model and the mathematical form of the interaction function. Our method is based on what we call \emph{decoupling approximation} for dynamical complex networks: we look for the best in-link configuration individually for each node by fitting the time series obtained by simulating each in-link configuration to the actual time series. This approach is much more efficient that the ideal method with exhaustive search, where one looks at \emph{all} link configurations for the entire network. Our approximation reduces the computational cost from $O(2^{N(N-1)})$ for the ideal method down to $O(N2^{N-1})$. As we have shown in detail, despite this trade-off our method displays excellent performance: it is extremely robust to the length and resolution of time series, it shows basically no dependence on the dynamical regime of the trajectories at hand (can extract information from relatively poor dynamics). Our method is also fairly robust to noise, even when on top of the noise we deal with short time series or time series with bad/low resolution. Therefore, the overall performance of our method is very close to the performance of the ideal (exhaustive) method, but with computational cost reduced by a factor of $O(2^N/N)$. We have also tested our method against dynamical networks of varying size and structure, revealing that method's performance does not depend critically on these parameters (except in terms of computational cost). In the reminder of the paper we discuss our findings, with particular emphasis on the current limitations of our method and prospects for its realistic employment.

We begin by discussing the range of validity of the decoupling approximation for dynamical networks. Actually, it is easy to see that our method cannot be immediately applied to non-directed networks, in which existence of link $j \rightarrow i$ implies the existence of the symmetric link $i \rightarrow j$. Assume that the statistics of in-link configurations for the node $i$ assign a high propensity for the link $j \rightarrow i$, but vice versa, the statistics for the node $j$ assigns a low propensity for the link $i \rightarrow j$. Our method, in its present form, does not include a way to reconcile these contradictory propensities, calling for a new intermediate algorithmic step able to reconcile them. On the other hand, we expect this approximation to be more valid in cases of directed networks in which shortest paths from a node to itself are longer, since that gets us further away from the non-directed case. In fact, any non-directed network can be seen as the limit case of a directed network, in which all links go both ways for all node pairs, which makes each node exactly two steps away from itself (assuming no self-loops). However, the successful employment of decoupling approximation for the purposes network reconstruction reported in this paper, make it a promising approach to study of the collective dynamics in complex systems. Also, given the ubiquity of directed networks in nature and society, this approach is likely to gain attention. Nevertheless, we recognize that a more complete investigation of the range of utility of decoupling approximation -- especially in relation to real complex systems -- remains a pressing research challenge.

Next we scrutinize our hypotheses, namely the knowledge of the dynamical model and of the interaction function. Indeed, in real complex systems neither is in general known, although in rare specific cases one or the other can be inferred with certain precision. Reconstruction methods that relax these hypotheses are a topic of intense ongoing research. However, we note that our method is able to reconstruct \emph{both} network topology \emph{and} network dynamics. This is in contrast to a large volume of literature related to methods that extract only the topology and disregard the dynamics. Yet in order to approach the reconstruction of network dynamics, one has to establish at least some general hypotheses, such as the form of the dynamical model. In fact, our approach can be easily modified to perform network reconstruction under any dynamical model by suitably changing the Equation~\ref{eq-1}. The ProBMoT framework that our method relies on allows for flexible specification of both dynamical model and interaction function $f$, but they both have to be known at least to some degree. Also, note that our method is immediately suitable also for scenarios where $f$ depends on the link, but again, all $f$s have to be known. This in principle extends its applicability to multiplex networks. This is also related to the problem of time series used for reconstruction. Longer time series will offer a better phase space coverage, while shorter time series may offer a better time resolution. The former yields more dynamical diversity and the latter more accurate derivative estimates. Of course, in any realistic scenario, experimental limitations are what "decides" what quality of data we have to work with. In other words, a good reconstruction method should be robust to both, but that can be so only when at least some reasonable hypotheses about general properties of the dynamical system are established.

The core question in the field of network reconstruction is the applicability of various methods to large real networks. We primarily envisage the application of our method in cases of networks where the details of interaction are known or can be reasonably well estimated. For instance, gene expression levels can nowadays be reliably measured for several genes simultaneously, although there are several hypotheses that one can make about the interaction functions. An additional limitation comes from the computational cost of $O(N2^{N-1})$, which still renders the application to large real networks computationally intractable. We note that this cost comes chiefly from looking at \emph{all} in-link configurations for a node, since their number, $2^{N-1}$, grows exponentially with network size. This calls for employment of adequate heuristics, which will reduce this cost via trade-off with reconstruction quality. For example, a version of simulated annealing could be used: one starts with any in-link configuration and computes the $\D$ associated with it. Then, a random mutation of in-link configuration is considered (e.g. changing one link for non-link or vice versa), and accepted (rejected) if it leads to better (worse) value of $\D$. Successively applying this procedure leads to (at least local) minimum of $\D$, but with far less mutation steps than $2^{N-1}$. While the computational cost is at present the main factor limiting the applicability of our method, we emphasize that this is a preliminary step towards a new class of reconstruction methods to be developed in future work along above described lines. Very important but seldom considered in the literature is the case of not having access to all network nodes (no total observability). This scenario is very realistic since a part of the network may not be accessible to empirical measurements. Here we have to rely on trajectories from some (but not all) nodes, and seek nevertheless to reconstruct the entire network (or as much of it as possible). There is a huge lack of suitable methods in this context and their development should be the subject of future work. Also in this context often comes the question of methods that instead of reconstructing the network topology, reconstruct the network model to which the reconstructed network belongs (e.g. scale-free or small-world). In other words, rather than reconstructing the network, one want to classify it in one of the classes defined by the shared topological characteristics. This remains an important open problem, deserving further work. 

We close the paper with discussion of how current network reconstruction approaches compare to one another. In fact, comparing reconstruction methods is not simple, since many of them start from different hypotheses and knowledge about the system, which makes their real merit harder to compare. The main distinction is between what different approaches are trying to reconstruct. While some approaches focus only on the network structure~\cite{roger,jure,tiago}, others (such as the one presented here) seek to reconstruct both structure and dynamics. Looking only at the network structure will in general give better results, but will also entirely neglect the dynamics. Certain methods give excellent results, but only when applied to cases of dynamics with specific properties, such as periodicity or synchronization~\cite{blaha,kralemann}. Another distinction runs along the ability to interfere with the system. While invasive methods give better results~\cite{us2}, interfering with the system is not always possible in practice. Almost all reconstruction methods rely on having some access to the observables in the system, yet some methods are more and some less robust to the quality of such dynamical data (e.g. time series)~\cite{timme1}. And finally, various methods make stronger or weaker assumptions on what is known about the dynamical network under study (e.g. interaction functions)~\cite{marc}. More assumptions in general lead to better performance, but information on whether the method's assumptions are satisfied is seldom available in real real systems. We hence conclude that the reconstruction concept presented in this paper is a promising new avenue in the field of network reconstruction, primarily due to its robustness to data quality. Still, the critical downside of our method is its computational cost, which scales as $O(N2^{N-1})$. Before its practical implementation becomes possible, future work will chiefly revolve around selecting the adequate heuristics to reduce this computational cost and re-examining the necessity of the hypotheses about knowledge of interaction functions.


\section*{Author contributions}

BŽ, ZL and LT envisaged the network reconstruction concept here presented and formulated the problem, NS and JT designed and carried out the empirical evaluation, NS and JT organized the results and prepared the figures, ZL, LT, NS and JT wrote the manuscript, all authors reviewed the manuscript. We consider the contributions of the first two authors, NS and JT, as equal. Similarly, we consider the contributions of the last three authors, ZL, LT and SD, as equal.


\begin{ack}
This research was supported by the Slovenian Research Agency (projects N2-0056, L2-7509, and V5-1657 and programs P1-0383, P2-0103 and P5-0093), by the Slovenian Ministry of Education, Science and Sport and the EU (grant C3330-17-529021), and by the EU (MSC-ITN-EJD grant COSMOS 642563, FP7-ICT-FET project MAESTRA 612944, H2020-FET-Flagship grant HBP SGA2 720270, and H2020-SC1-RIA project SAAM 769661).
\end{ack}

\section*{References}
\bibliographystyle{unsrt}

\newpage
\appendix

\section{More on algorithmic aspects of our reconstruction method} 

As stated in the main text, our reconstruction method is based on \textit{ProBMoT}, which is a computational framework for equation discovery that allows to infer laws governing certain dataset, whether the laws are physical or otherwise. In this Appendix we further describe ProBMoT and its general functioning, which serves to better understand the full extent of our reconstruction method and its background.

ProBMoT implements the process-based approach to automated modeling of dynamical systems. Process-based modeling is a recent development within the long tradition of work in equation discovery: it combines domain-specific modeling knowledge with actual time-series data. It simultaneously addresses the tasks of identifying the mathematical form of the equations and their parameters estimation, making exhaustive use of the numerical simulations. In particular, ProBMoT employs constrained enumeration to explore the search space of all possible model equations, which is implicitly defined by modeling knowledge provided at input. In addition, ProBMoT uses time-series data, also provided at input, to estimate the values of the parameters for each potential model equation. Next, ProBMoT measures the fit between the models and the data as the discrepancy between the simulation of the model and the input data. As a result, ProBMoT outputs a list of possible model equations, ranked according to the fit of the model to the input data in descending order (best fit first). From this ranking, one can establish a plateau of best models and extract the single best solution.

What we have discussed in the Reconstruction method Section, is nothing but a specific case of the above described general procedure. For the particular task of reconstructing an in-link configuration for a single network node $i$, we encode two components of domain-specific modeling knowledge. The first, depicted in Table~\ref{tab:probmot-lib}, encodes general knowledge for establishing a process-based model of dynamics of an arbitrary network. 
\begin{table}[!ht]
	\caption{\label{tab:probmot-lib} ProBMoT specification of template entities and processes for process-based modeling of network dynamics. See text for details.}
	\begin{indented}
	\ttfamily 
	
	\item[]\begin{tabular}{>{\footnotesize  }l}
	\br
		template entity \textsl{Node} \{ \\
		\qquad vars: \textsl{value} \{aggregation:\textsl{sum};\} \} \\
		template process \textsc{Edge} (\textsl{s}:\textsl{Node}, \textsl{d}:\textsl{Node}) \{\}\\
		\qquad template process \textsc{thEdge}:\texttt{\textsc{Edge}} \{ \\
		 \qquad \qquad equations: $ \mathrm{td(d.value) =tanh(s.value)} $; \}\\
		\qquad template process \textsc{noEdge}:\texttt{\textsc{Edge}} \{ \\
		\qquad \qquad equations: $ \mathrm{td(d.value) =0} $; \}\\
	\br
	\end{tabular}%
	\end{indented}
\end{table}%
Here, network nodes are represented as model entities (template entity {\ttfamily\textsl{Node}}), while the edges are represented as processes of interactions between pairs of network nodes (template process {\ttfamily\textsl{Edge}}). The template process representing the edges has two modeling alternatives: one models absence of edge between the particular pair of nodes $s$ and $d$, i.e., $ A_{sd} = 0 $, while the other models its presence, i.e., $A_{sd}=1$. Note that the latter alternative encodes the mathematical form of the function $f$ that models the way the nodes interact among them ($ f = \tanh $). Other functional forms can be considered by adding additional modeling alternatives to the {\ttfamily\textsl{Edge}} template process.

The second component of the knowledge for process based modeling, depicted in Table~\ref{tab:probmot-task}, specifies the particular task of modeling the in-link configuration of a given network node $i$. 
\begin{table}[!ht]
	\caption{\label{tab:probmot-task} ProBMoT specification of the task of modeling the in-link configuration of a network node $i$. See text for details.}
	\begin{indented}
	\ttfamily 
	\item[]\begin{tabular}{>{\footnotesize  }l}
	\br
		entity $\mathtt{node_{i}}$:\textsl{Node} \{vars:\textsl{value}\{role:endogenous;\} \} \\
    	j = 1..$N$, j $ \neq$ i:\\
		 \qquad entity $\mathtt{node_{j}}$:\textsl{Node} \{vars:\textsl{value}\{role:exogenous;\} \}\\
		 \qquad process \textsc{edge}$\mathtt{_{ji}}$($\mathtt{node_{j}, node_{i}}$):\textsc{Edge}\{\};\\
	\br
	\end{tabular}%
	\end{indented}
\end{table}%
The latter is encoded as an endogenous (internal) entity, the dynamics of which is being modeled (a system composed of endogenous entities only is an autonomous system). All the other nodes $ j \neq i $ are declared as exogenous (external) entities. Their dynamics is not being modeled and they can only appear as input on the right-hand side of the equation modeling the dynamics of $i$. This sophisticated organization into templates and entities serves to facilitate the employment of ProBMoT in diverse scenarios, above described being one of them.

Given this specification and the mathematical forms encoded in the library of templates and entities from Table~\ref{tab:probmot-lib}, ProBMoT first enumerates the $2^{N-1}$ candidate in-link configurations. It then simulates each and ranks them with respect to increasing error $\D$, defined in Equation~\ref{eqrmse}. For illustration, an example ProBMoT model representing the correct in-link configuration of node $4$ in the network in Figure~\ref{fig-network} is given in Table~\ref{tab:probmot-model}: The two links from nodes $6$ and $7$ to node $4$ correspond to the two {\ttfamily\textsl{thEdge}} processes, while the absence of links from all other nodes to node $4$ is represented by the 17 {\ttfamily\textsl{noEdge}} processes. 
\begin{table}[!ht]
	\caption{\label{tab:probmot-model} An example ProBMoT model of the in-link configuration of network node $4$ in Figure~\ref{fig-network}. See text for details.}
	\begin{indented}
	\ttfamily 
	\item[]\begin{tabular}{>{\footnotesize  }l}
	\br
		entity $\mathtt{node_{4}}$:\textsl{Node} \{vars:\textsl{value}\{role:endogenous; initial:-6.25\} \} \\
        entity $\mathtt{node_{1}}$:\textsl{Node} \{vars:\textsl{value}\{role:exogenous;\} \}\\
        entity $\mathtt{node_{2}}$:\textsl{Node} \{vars:\textsl{value}\{role:exogenous;\} \}\\
        entity $\mathtt{node_{3}}$:\textsl{Node} \{vars:\textsl{value}\{role:exogenous;\} \}\\
        entity $\mathtt{node_{5}}$:\textsl{Node} \{vars:\textsl{value}\{role:exogenous;\} \}\\
        entity $\mathtt{node_{6}}$:\textsl{Node} \{vars:\textsl{value}\{role:exogenous;\} \}\\
        entity $\mathtt{node_{7}}$:\textsl{Node} \{vars:\textsl{value}\{role:exogenous;\} \}\\
        \vdots \\
        entity $\mathtt{node_{20}}$:\textsl{Node} \{vars:\textsl{value}\{role:exogenous;\} \}\\
        \\
	    process \textsc{edge}$\mathtt{_{14}}$($\mathtt{node_{1}, node_{4}}$):\textsc{noEdge}\{\};\\
	    \vdots \\
	    process \textsc{edge}$\mathtt{_{54}}$($\mathtt{node_{5}, node_{4}}$):\textsc{noEdge}\{\};\\
	    process \textsc{edge}$\mathtt{_{64}}$($\mathtt{node_{6}, node_{4}}$):\textsc{thEdge}\{\};\\
	    process \textsc{edge}$\mathtt{_{74}}$($\mathtt{node_{7}, node_{4}}$):\textsc{thEdge}\{\};\\
	    process \textsc{edge}$\mathtt{_{84}}$($\mathtt{node_{8}, node_{4}}$):\textsc{noEdge}\{\};\\
	    \vdots \\
	    process \textsc{edge}$\mathtt{_{204}}$($\mathtt{node_{20}, node_{4}}$):\textsc{noEdge}\{\};\\
	\br
	\end{tabular}%
	\end{indented}
\end{table}%

While in this paper we use exhaustive search of the space of candidate in-link configurations, in future work we intend to use incomplete, heuristic search strategies to look for the optimal in-link configurations. The clarification in this Appendix also serves to understand that switching from exhaustive search to heuristic search can be done relatively simply within ProBMoT framework. The next question here revolves around selection of the best heuristic search strategy that leads to the best trade-off between reconstruction quality and computational cost. We will investigate this important question in future work, as it is vital for application of our method to large real networks.


\section{Choice of trajectories}

We use this Appendix to further substantiate our choice of five trajectories used for testing the performance of our reconstruction method. As stated in the main text, we want to cover as much dynamical diversity as possible in order to test our method against different dynamical regimes, i.e., varying richness of dynamics. However, our interest is also to test the method against noisy time series of variable length and resolution, since real scenarios almost never involve noiseless time-resolved data of arbitrary length and resolution. This unfortunately hinders the precise quantification of the dynamical diversity via standard non-linear dynamics tools, since they require longer time series of fixed resolution to yield interpretable measures. Even if we were to use for example Fuzzy of Shannon entropy to quantify the information content of trajectories, another problem arises due to each time series (each node) having its own value of entropy. Simply averaging these values for all nodes loses some of the information and makes it harder to tell rich dynamics from poor. We thus give a qualitative argument to support our choice of five trajectories. To this end we make the following two observations.

\begin{figure*}[!htb]
\includegraphics[width=\textwidth]{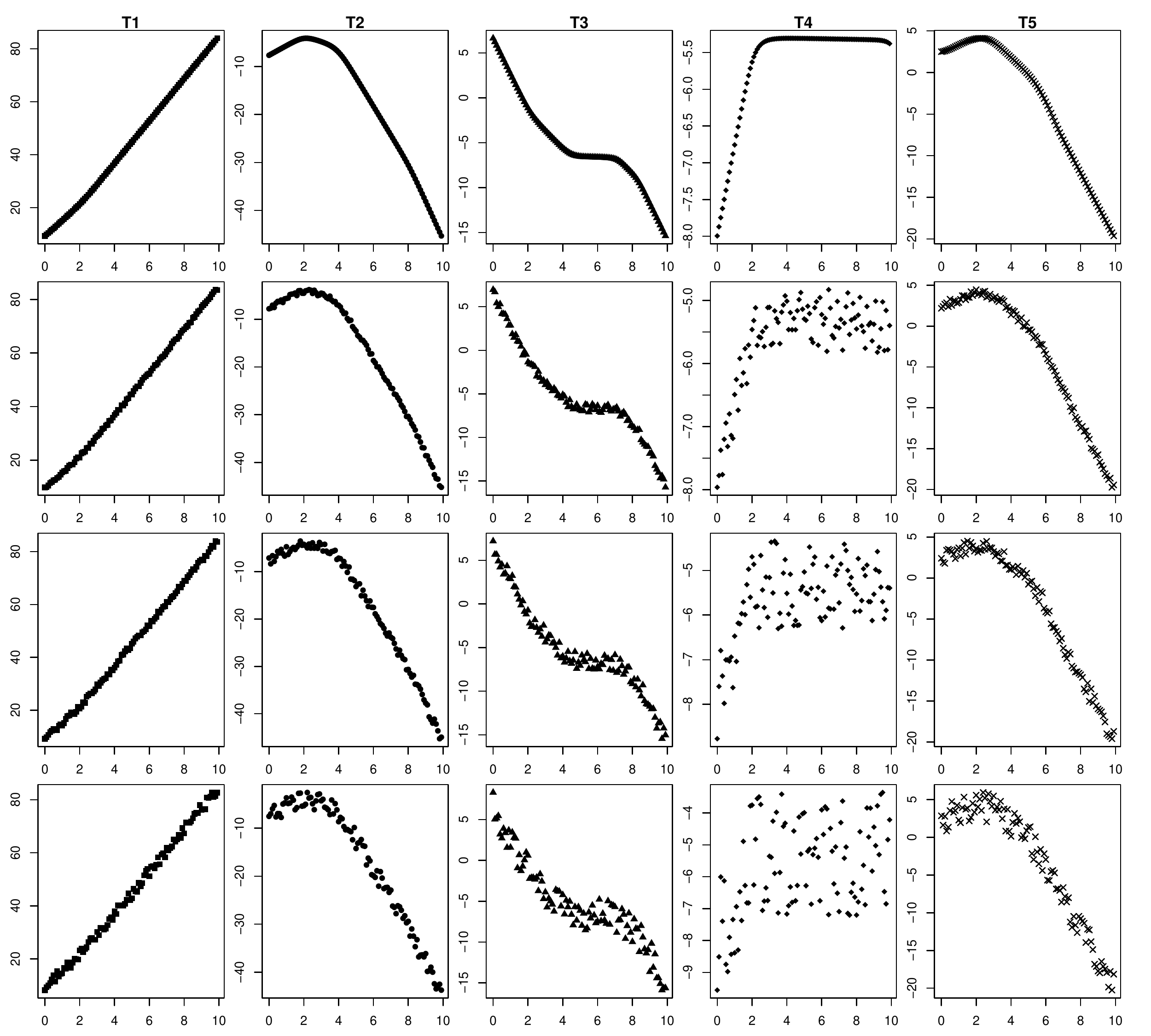}
\caption{Time series for the node 7 ($x_7 (t)$) for all five selected trajectories $\mathbf{T}_1$,... $\mathbf{T}_5$. Each column corresponds to one trajectory, as indicated in the title. Top panel: noise-free case $\eta=0$. Second, third and bottom panel, noise strength $\eta=0.5$, $\eta=1.0$ and $\eta=2.0$, respectively. Note that for clarity the range in $y$ axis is adjusted to the range of values of $x_7 (t)$ for each case separately. Node 7 is selected since it is fairly representative of all other nodes for each specific trajectory. In fact, node 7 is very central in the examined network (cf. Figure~\ref{fig-network}).}
\label{figtrajectories}
\end{figure*}

First, we visually illustrate the diversity of the dynamics by plotting the time series for node 7 ($x_7 (t)$) for all five trajectories in Figure~\ref{figtrajectories} (top panel). We pick node 7 since it most illustratively resembles the dynamics of other nodes for that trajectory. Clearly, the variability of time series differs from trajectory to trajectory. We also note that in fact, none of the time series is extremely rich in its dynamical behavior.
Next, for better orientation in the main text, we also show the time series for the same node, but for all three considered noise strengths: $\eta=0.5$ (second panel), $\eta=1.0$ (third panel), $\eta=2.0$ (bottom panel). In particular, we emphasize that third noise level is in fact very strong with respect to the range of time series variability.

\begin{figure}
\centering
\includegraphics[width=.5\textwidth]{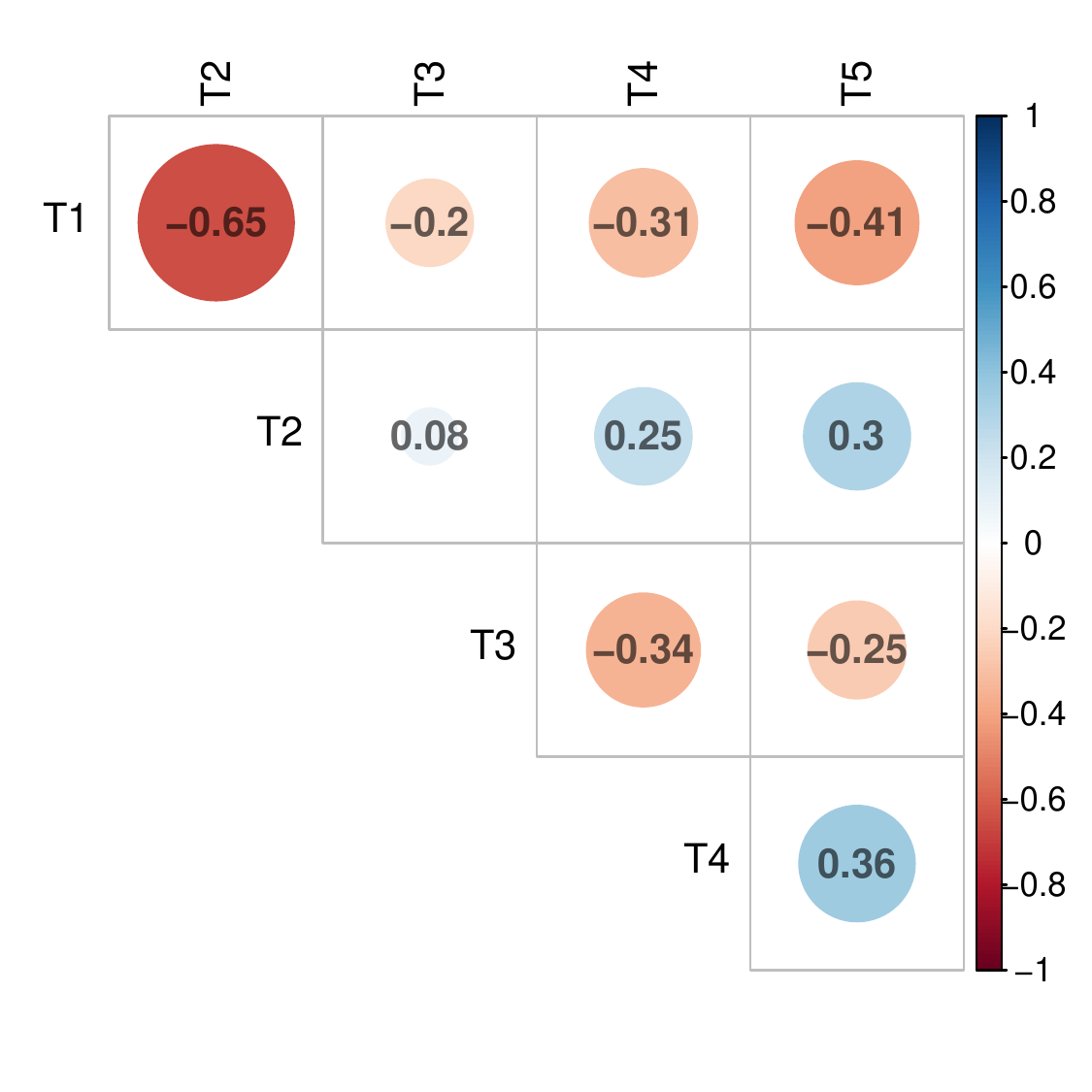}
\caption{Matrix of pairwise Pearson correlation coefficients for the five trajectories used for network reconstruction. The correlation coefficient is calculated by first concatenating 20 time series for one trajectory into a single sequence and then calculating the respective coefficients pair-wise.}
\label{correlations}
\end{figure}

Second, we compute the standard Pearson correlation between pairs of different trajectories. To that end we concatenate all time series for each trajectory and compute the Pearson correlation coefficient for each pair. Results are reported in Figure~\ref{correlations}. 
Indeed, none of the pairs is strongly correlated. This indicates that the considered trajectories are fairly independent (at least) pair-wise and display different uncorrelated dynamical behaviors. This supports our argument that this choice of trajectories represents, at least to some degree, the dynamical patterns exhibited by the considered dynamical network. Also, it suggests that method's performance from one of the trajectories cannot be immediately related to performance from other trajectories. In other words, by testing our method on these five trajectories we indeed make five independent tests. We argue that above two observations provide enough evidence to demonstrate, at least in principle, that our five trajectories indeed reflect qualitatively different dynamical regimes.

\end{document}